\documentclass[sigconf,manuscript]{acmart}

\usepackage{graphicx}
\usepackage{booktabs}
\usepackage{multirow}
\usepackage{paralist}
\usepackage{url}
\usepackage{hyperref}

\acmConference[LLM4Eval '25]{Workshop on Large Language Models for IR Evaluation}{February 2025}{Virtual Event}



\begin{document}

\title{Validating LLM-Generated Relevance Labels for Educational Resource Search}
\author{Ratan J. Sebastian}
\email{Ratan.Sebastian@tib.eu}
\orcid{0000-0002-8034-3382}
\affiliation{
  \institution{L3S Research Center, Leibniz University Hannover}
  \city{Hannover}
  \country{Germany}
}
\author{Anett Hoppe}
\email{anett.hoppe@tib.eu}
\orcid{0000-0002-1452-9509}
\affiliation{
  \institution{TIB -- Leibniz Information Centre for Science and Technology}
  \city{Hannover}
  \country{Germany}
}
\affiliation{
  \institution{L3S Research Center, Leibniz University Hannover}
  \city{Hannover}
  \country{Germany}
}

\renewcommand{\shortauthors}{Sebastian, et al.}

\begin{abstract}
Manual relevance judgements in Information Retrieval are costly and require expertise, driving interest in using Large Language Models (LLMs) for automatic assessment.
While LLMs have shown promise in general web search scenarios, their effectiveness for evaluating domain-specific search results, such as educational resources, remains unexplored.
To investigate different ways of including domain-specific criteria in LLM prompts for relevance judgement, we collected and released a dataset of 401 human relevance judgements from a user study involving teaching professionals performing search tasks related to lesson planning.
We compared three approaches to structuring these prompts:
a simple two-aspect evaluation baseline from prior work on using LLMs as relevance judges, 
a  comprehensive 12-dimensional rubric derived from educational literature, 
and criteria directly informed by the study participants.
Using domain-specific frameworks, LLMs achieved strong agreement with human judgements (Cohen's $\kappa$ up to 0.650), significantly outperforming the baseline approach.
The participant-derived framework proved particularly robust, with GPT-3.5 achieving $\kappa$ scores of 0.639 and 0.613 for 10-dimension and 5-dimension versions respectively.
System-level evaluation showed that LLM judgements reliably identified top-performing retrieval approaches (RBO scores 0.71-0.76) while maintaining reasonable discrimination between systems (RBO 0.52-0.56).
These findings suggest that LLMs can effectively evaluate educational resources when prompted with domain-specific criteria, though performance varies with framework complexity and input structure.

\end{abstract}

\maketitle

\section{Introduction}
Large Language Models (LLMs) are increasingly being explored as a way to address the resource-intensive challenge of generating relevance judgements for Information Retrieval (IR) evaluation~\cite{faggioli}.
While recent research suggests LLMs can achieve human-comparable performance in general relevance judgements for web-searches~\cite{thomas}, their reliability for domain-specific, multi-dimensional relevance judgements remains an open question.
This is particularly crucial in professional search contexts like education, where relevance assessment requires evaluating resources across multiple domain-specific criteria that reflect both enduring professional needs (grade level appropriateness, pedagogical fit, etc.) and immediate task requirements (learning resource type, current class topic, etc.).

Prior research suggests that while many professional searchers rely on specialised repositories~\cite{russell}, teachers predominantly use general web search~\cite{cortinovis,de_los_arcos,de_medio}.
This requires them to layer their pedagogical expertise onto standard web search relevance criteria.
They must evaluate resources across numerous dimensions including content match, grade-level appropriateness, student engagement, and pedagogical soundness~\cite{reitsma,diekema,wetzler,bortoluzzi}. 

This paper investigates how effectively LLMs can generate reliable relevance judgements for educational resource search while accounting for these domain-specific complexities.
Previous research into using LLMs in the zero-shot setting~\cite{thomas,faggioli} has found effective prompting strategies for relevance judgement of general web search documents.
Particularly, Thomas et al.~\cite{thomas} showed that the most important component of a prompt for relevance judgements were the "aspects" of the judgement;
i.e. the sub-dimensions of the overall judgement on which the LLM was asked to score the resource first before aggregating into an overall score.
In this work, we examine how well this strategy works for domain-specific, professional relevance judgements in the context of educational searches by teachers.
We specifically examine two key research questions:

\begin{itemize}
    \item \textbf{RQ1}: Which combination of relevance dimensions leads to the best agreement with the relevance judgements of human searchers?
    \item \textbf{RQ2}: Do LLM relevance judgements lead to similar query and system rankings as human judgements? 
\end{itemize}

To address these questions, we used data from a user study where participants performed typical lesson planning-related tasks. 
During the study, teachers were asked to think aloud about their relevance judgements.
These transcripts were coded on the standard TREC relevance judgement scale and used as ground truth~\cite{TRECRobust2004}.
We derived various breakdowns of relevance dimensions from the literature on teacher search~\cite{reitsma,wetzler} as well as from the results of the user study itself.
To guide LLMs in making relevance assessments, we incorporated these education-specific relevance dimensions into the best-performing prompt template from prior work on LLM-based IR relevance assessment~\cite{thomas}. 
We then compared how well the LLM performed when using these domain-specific dimensions versus the more generic aspects from this prior work: trustworthiness and content match.

Through this investigation, we contribute to both the practical implementation of LLM-based IR evaluation and the theoretical understanding of how automated systems can support human judgement in professional search contexts.
Our findings provide insight into which aspects of educational resource relevance are important for LLMs to make good overall relevance judgements and which require more careful human validation
Our methodology offers a practical framework for implementing LLM-based relevance assessment in complex professional domains.

\section{Related Work}
In this section, we survey important related work in the fields of relevance judgements broadly for professional search and then specifically for educational resources. We then cover recent work in using LLMs for relevance judgements and the challenges of making these judgements in domain-specific contexts.

\subsection{Relevance in Professional Search}

Professional search encompasses both specialised database search and general web search adapted for domain-specific needs~\cite{russell,cortinovis}.
Across domains like healthcare, law, patents and education, professionals must evaluate resources against multiple domain-specific criteria that reflect both enduring professional contexts and immediate task requirements~\cite{russell,reitsma}.
For instance, healthcare professionals evaluate resources against regulatory requirements and evidence standards, while legal researchers consider jurisdictional constraints and case currency~\cite{russell}.

Teachers represent an interesting case within professional search, as they primarily use general web search tools rather than specialised repositories~\cite{de_medio,richters}.
They must layer their pedagogical content knowledge  onto general search interfaces to evaluate resources across multiple dimensions including content match, situational fit, affective match~\cite{reitsma} and reputation judgements~\cite{wetzler}.

\subsection{Relevance in Educational Resources}

The evaluation of educational resources introduces unique challenges due to the multi-dimensional nature of pedagogical relevance.
Reitsma et al.~\cite{reitsma} identified nine specific alignment criteria for K-12 teaching, including student motivation, concept coverage, grade-level appropriateness, and the availability of supplementary materials.
Their research demonstrated that breaking down relevance into specific dimensions produces more reliable judgements from human annotators compared to overall alignment assessments.

Teachers' evaluation strategies are characterised by partial rather than holistic assessment~\cite{diekema}.
Rather than evaluating documents as complete units, teachers focus on how specific parts can be adapted to their needs.
This aligns with Bortoluzzi et al.'s~\cite{bortoluzzi} findings that teachers prefer finding adaptable authentic materials over pre-made educational resources, viewing the search process itself as an opportunity for professional development.
Both of these findings speak to the complexity and shifting nature of teacher search and how various dimensions might become more or less important based on search task.
Any approach to building a test collection for teacher search must take into account the need to evaluate documents exhaustively and the need for judgements under varying task conditions.
This makes building such a collection a more resource-intensive task than for general web search.

\subsection{LLMs for Relevance Assessment}

Recent work has demonstrated that Large Language Models can match or exceed human performance in various assessment tasks.
Thomas et al.~\cite{thomas} showed that LLMs could achieve agreement levels with human assessors comparable to the inter-annotator agreement between trained professionals, with Cohen's $\kappa$ ranging from 0.50 to 0.72 depending on prompt design.

However, LLM performance varies considerably based on prompt design and even simple paraphrasing~\cite{thomas}.
This sensitivity to prompting raises important questions for educational resource evaluation, where the assessment criteria are complex and multifaceted.
The success of LLMs appears to depend heavily on carefully structured prompts that break down the evaluation task into explicit steps or aspects.
In this work, we try to find out which aspects are important for this domain-specific relevance judgement of educational resources by teachers.

\subsection{Challenges in Domain-Specific Relevance}

Given this reliance of LLM evaluation on appropriate aspects of a relevance judgement, the question becomes: "Which aspects?"

Prior work on teacher search has suggested that teachers search for a variety of reasons with shifting relevance requirements based on their context~\cite{bortoluzzi}.
For instance, the relevance requirements during exploratory searches for teaching ideas might be different than those for finding resources for use in a classroom.
The system must be able to adjust the relative weights of these aspects based on the search task.

There is a rich literature based on surveys and interviews with teachers that give us an idea about the contexts in which teachers search~\cite{bortoluzzi,de_los_arcos}, how they search~\cite{yalcinalp,diekema} and how they judge resources encountered during that search~\cite{wetzler,reitsma}.
Specifically of interest to this study is the work by Reitsma et al.~\cite{reitsma} which sought to bring some objectivity to the very subjective notion of relevance as defined by Saracevic~\cite{saracevic}.
They identified several relevance dimensions that were more objectively measurable and led to higher inter-annotator agreement than the overall relevance.
However, the dimensions they identified do not cover all that are indicated by the qualitative work on the subject.
For instance, De Los Arcos et al.~\cite{de_los_arcos} found that the "resource being created by a reputable person or institution" was an important factor in resource selection.
Yalcinalp et al.~\cite{yalcinalp} found that teachers valued being able to find resources by learning goals.
In work about the related concept of learning resource quality, Wetzler at al.~\cite{wetzler} identified dimensions of quality that cover these aspects of learning resources and showed, like~\cite{reitsma}, that they were more objectively measurable than overall quality.
We employ a combination of both these models to capture as many components of learning resource relevance as possible.

\section{Methodology}
Our investigation of LLM-based relevance assessment for educational resources followed three main steps. First, we collected ground-truth data through a user study where participants evaluated resources, collecting think-aloud protocols. This allowed us to capture relevance judgements and explicit assessment criteria. Second, we designed and implemented different frameworks for LLM-based assessment, exploring various ways to structure the relevance judgement task. Finally, we evaluated the effectiveness of these approaches both at the document level and for system-level evaluation tasks. 
\subsection{Ground Truth Collection}
 
\subsubsection{User Study Design}
We conducted a user study with 12 participants representing different levels of teaching expertise: pre-service teachers (4), in-service teachers (3), and educational researchers with teacher education experience (5).
All participants had extensive familiarity with lesson planning for grades 5-10 through teaching, studying, or professional practice.

Tasks were personalised for each participant based on their subject expertise. This was done to ameliorate the effects of prior knowledge that was seen to be an issue in a prior pilot study (i.e., search tasks which reduced to known-item lookup for teachers who were very familiar with a topic).
Prior to task generation, participants rated different curriculum topics on a 5-point scale across three dimensions: conceptual familiarity, teaching experience, and familiarity with online resources.
Topics were selected where participants reported high conceptual familiarity (4-5/5) but lower teaching and resource familiarity (1-3/5).

\subsubsection{Data Collection}

Participants were instructed to think aloud while they performed their search tasks.
Utterances were recorded along with timestamps.
Their interactions with the browser were also recorded yielding a log of the documents they visited and which document they were looking at at which time.
For the think-aloud, participants were instructed to focus on verbalising their relevance criteria and decision-making processes.
They were prompted by the study conductor if they forgot to verbalise.
The timestamped transcriptions were merged with the search log to identify which document they were looking at when they were talking.
Participants completed 6 search tasks over the course of about 90 minutes during which they made an average of 5.4 queries and visited 9.1 documents per task.

\subsubsection{Transcription Coding}
From these transcriptions, we identified and extracted all statements related to document relevance judgements.
The statements were coded into the categories: Search Strategy Statements (SST), Content Assessment (CON), Navigation Decisions (NAV), Interface Commentary (INT), Task Management (TSK) and Subject Matter Knowledge (SMK).
Of these, the Content Assessment statements were used in this study.
Content Assessment statements were verbal expressions where the participant assesses the quality, appropriateness, credibility, or usefulness of content or its source. This includes evaluating the authority of sources, assessing whether the content matches curricular needs, checking if the material is at an appropriate level, and determining if the content is accurate and complete.
Some example statements include:
\begin{itemize}
    \item "from the University of Göttingen, that sounds serious and didactically sound to me at first." - evaluating source credibility.
    \item "that's already a bit too complicated for 7th, 8th grade, but you could use this part here" - evaluating pedagogical appropriateness.
\end{itemize}
The statements were then coded by the author according to the TREC standard of 2 = highly relevant, 1 = partly relevant and 0 = not relevant~\cite{TRECRobust2004}.
"Not Relevant" statements were \begin{inparaenum}
    \item quick dismissive judgements without explanation (e.g., "I don't like this") or
    \item evaluations of content/sources as unrelated to the search task
\end{inparaenum}.
"Somewhat Relevant" statements were \begin{inparaenum}
    \item partial judgements of the document in relation to the task,
    \item statements about aspects of usefulness or appropriateness or
    \item preliminary judgements that required further evaluation
\end{inparaenum}.
"Highly Relevant" statements were \begin{inparaenum}
    \item identifying documents as specifically aligned with task requirements
    \item the final statement of a search session that ended successfully
\end{inparaenum}.
If there were multiple statements for a single document they were judged altogether to take the final meaning of the participant.
There were a total of 72 search tasks across 12 participants but only utterances for 19 tasks from 7 participants were included in the final dataset as being unambiguous.
The final dataset includes 401 relevance judgements across 353 documents for 19 topics.
This dataset is made available at \url{https://tinyurl.com/er-rel-data}.

\subsubsection{Document Pre-processing}
Since the documents were web pages or PDF documents, they needed to be pre-processed before being used for automated LLM judgements.
The input modalities tested were text and image.
Text was extracted from both documents using a standard parser.
Images of HTML documents were taken using screen captures of the rendered page in a browser.
Images of PDF documents were constructed by concatenating images of each page.

\subsection{LLM-Based Assessment Design}
Our LLM-based assessment approach addressed two key design challenges: how to present document content to the LLMs within context window constraints, and how to structure the prompts to capture domain-specific relevance criteria.
\subsubsection{Input Processing}
To investigate how different amounts of document content affect LLM judgements, we developed two document representations to include in the prompt.
\begin{enumerate}
    \item First-impression representation (\textbf{HEAD}): Initial 5,000 characters (approximately 2 pages), simulating quick relevance judgements made from a document's beginning
    \item Comprehensive representation (\textbf{SKIM}): Ten 1,000-character segments sampled from the first 50,000 characters (approximately 20 pages), simulating how readers skim through longer documents
\end{enumerate}

The limit of 50,000 was selected to fit into the context windows of all evaluated LLMs. This covered most of our corpus, with only 1.2\% of documents exceeding this limit (though 50.8\% were longer than the 5,000-character HEAD representation).

\subsubsection{Prompt Exploration Framework}
\label{sec:frameworks}
Building on Thomas et al.'s finding that "DNA" prompts (Description, Narrative, and Aspects) perform best for relevance assessment, we focused on exploring different aspect frameworks for educational resource evaluation.
We developed and compared five frameworks:

\paragraph{Baseline Framework (M0):} As a baseline we selected Thomas et al.'s~\cite{thomas} two aspects: trustworthiness and content match.
We call this model \textbf{M0}.

\paragraph{Literature-Derived Frameworks (M1\&2):} We then used a framework derived from Reitsma et al.~\cite{reitsma} and Wetzler et al.~\cite{wetzler} who looking into dimensions of educational resource relevance and the related concept of resource quality.
From these, we assembled the 12-dimensional resource relevance framework shown in Table \ref{tab:relevance_model_hierarchical}(D1.1 - D5.1).
We will call this model \textbf{M1}.
We further grouped these dimensions into higher-level groups (D1-D5) to examine the effectiveness of a smaller number of aspects that still measure the same educational dimensions.
This will be called \textbf{M2}.

\begin{table*}
\caption{Hierarchical dimensions of the relevance model (M1 \& M2) from Reitsma et al.~\cite{reitsma} and Wetzler et. al~\cite{wetzler}. These are shorted descriptions, full descriptions that are used for LLM prompts are at \url{https://tinyurl.com/relevance-gap-dimensions}}
\label{tab:relevance_model_hierarchical}
\begin{tabular}{p{0.8cm}p{0.8cm}p{3.2cm}p{11cm}}
\toprule
\textbf{M1 - ID} & \textbf{M2 - ID} & \textbf{Dimension} & \textbf{Description} \\
\midrule
D1 &     & Affective Match & Document is intended and designed for educational use \\
     & D1.1 & Motivation & Materials that are motivating or stimulating to students \\
     & D1.2 & Identifies Learning Goals & Mentions the knowledge and skills a student is expected to acquire \\
     & D1.3 & Organised for Learning Goals & The document is structured according to its learning goals \\

\midrule
D2 &     & Content Match & Document is relevant to the school-subject specific topics needed by the teacher \\
     & D2.1 & Concepts & Contains keywords/terms from the information need \\
     & D2.2 & Background & Provides relevant background material \\
\midrule
D3 &     & Object Match & Document is relevant to the teachers' teaching context \\

   & D3.1 & Grade level & Is appropriate for the grade level in the information need \\
\midrule
D4 &     & Situational Match & Document is of the relevant resource type that the teacher is looking for \\
     & D4.1 & Non-textuals & Has non-textual items pertaining to the information need \\
     & D4.2 & Examples & Has real-world examples pertaining to the information need \\
     & D4.3 & Hands-on & Has hands-on activities pertaining to the information need \\
     & D4.4 & Attachments & Has attachments; e.g. score sheets, rubrics, test questions, etc. \\
     & D4.5 & References & Has references or internet links to relevant material elsewhere \\
\midrule

D5 &      & Reputation & Document comes from a reputable source \\
   & D5.1 & Prestigious Publisher & The document's publisher is recognisable and is prestigious and trustworthy \\
\bottomrule
\end{tabular}
\end{table*}

\paragraph{User-Derived Framework (M3):} We then tried to use our findings from the user study to see if the relevance criteria that the participants themselves identified would be a better aspect framework for the LLM prompt.
These participant dimensions (\textbf{M3}) are shown in Table \ref{tab:participant_dimensions}. We evaluated both the top 10 (\textbf{M3(10)}) and top 5 (\textbf{M3(5)}) dimensions in our prompts.

\begin{table*}
\caption{Top 10 participant-identified dimensions for evaluating educational resources}
\label{tab:participant_dimensions}
\begin{tabular}{p{0.8cm}p{4.5cm}p{10.3cm}}
\toprule
\textbf{M3 - ID} & \textbf{Dimension} & \textbf{Description} \\
\midrule
PD1 & Source Credibility & Document comes from trusted educational sources (e.g., educational institutions, recognized journals, state education servers) with clear pedagogical expertise \\
\midrule
PD2 & Content Quality and Accuracy & Document demonstrates subject matter expertise with accurate content, proper terminology, and clear explanations/justifications \\
\midrule
PD3 & Grade-Level Appropriateness & Content matches the target grade level in terms of difficulty, language, and complexity \\
\midrule
PD4 & Complete Teaching Materials & Document includes comprehensive materials (worksheets, answer keys, teacher notes) with clear implementation instructions \\
\midrule
PD5 & Accessibility & Document is freely available without paywalls or registration requirements \\
\midrule
PD6 & Adaptability & Content can be modified for specific classroom needs and different pedagogical approaches \\
\midrule
PD7 & Implementation Feasibility & Document works within time constraints and requires only available resources/equipment \\
\midrule
PD8 & Student Engagement & Document promotes active learning through interactive elements and allows students to make their own observations \\
\midrule
PD9 & Differentiation Support & Materials include scaffolding and options for different learning speeds and approaches \\
\midrule
PD10 & Curriculum Integration & Content aligns with curriculum requirements and builds upon standard approaches \\
\bottomrule
\end{tabular}
\end{table*}

\subsection{Evaluation Methodology}
We evaluated the effectiveness of LLM-generated relevance judgements at three levels: agreement with human judgements at the document level, impact on system evaluation, and contribution of individual dimensions to the overall assessments.
\subsubsection{Document-Level Agreement}
To measure agreement between LLM and human judgements at the document level, we use Cohen's $\kappa$. 
Following previous work in relevance assessment~\cite{thomas, faggioli}, we consider this both with the original three-level judgements (0-2) and with binary relevance where partially and highly relevant documents are conflated.

\subsubsection{System Evaluation Impact}
To understand how well LLM-generated labels serve for evaluating retrieval systems, we examine two key aspects as in Thomas et al.~\cite{thomas}:

\textbf{Identifying difficult queries:} We assessed whether LLMs identify the same challenging queries as humans by ranking queries based on mean system performance. Rankings were compared using rank-biased overlap (RBO) with $\phi=0.9$, corresponding to examining approximately the worst 10 queries.

\textbf{Distinguishing system performance:} We tested whether LLM judgements could identify top-performing systems similarly to human judgements.
Rankings were compared using RBO with $\phi=0.7$, focusing on the top 3-4 systems.
The retrieval systems were specifically selected to include multiple variants of different retrieval families with controlled parameter changes -- BM25 with $b$ values of 0.3 and 0.7 and $k1$ of 1.5, TF\_IDF with and without normalisation, DirichletLM with $\mu$ values of 500 and 2000, and PL2 with $c$ values of 5 and 10.
This systematic variation allowed us to test whether human and LLM judgements can detect meaningful differences between similar systems, requiring more fine-grained relevance assessments than simply distinguishing between fundamentally different approaches.

In all cases, RBO scores are normalised so that 0 represents an exactly opposite ranking and 1 represents identical rankings as in~\cite{thomas}.
We computed system performance using standard IR metrics including Mean Average Precision (MAP), Precision@10 (P@10), and normalised Discounted Cumulative Gain (nDCG@10).

\subsection{Dimensional Contribution Analysis}

To analyse the relative importance of different dimensions in each relevance assessment framework, we calculated correlations between individual dimension scores and overall relevance judgements across all assessed documents.
The squared correlation coefficients were used to estimate the percentage of variance in overall relevance judgements explained by each dimension.
For each framework, we identified the three dimensions with the highest explanatory power and summed their contributions to obtain an upper bound on their combined explanatory power.
This analysis was conducted using both correlation analysis and linear regression to validate the importance of dimensions, with correlation coefficients being used for the final variance calculations due to their more straightforward interpretation in terms of explained variance.

\section{Results}
In this section, we discuss the results of evaluating LLM relevance judgements on the basis of their direct agreement with human judgements and their indirect agreement with system-level evaluations produced by human judgements. We conclude with the results of the dimensional contribution analysis. 

\subsection{Agreement Analysis}
\begin{table}[t]
\caption{Cohen's $\kappa$ comparison across different models and evaluation metrics. Bold values indicate scores > 0.6. Models: GPT = GPT-3.5-turbo-0125 (text) \& gpt-4-mini-2024-07-18 (images); Llama = meta-llama/Llama-3.3-70B-Instruct (text) \& Llama-3.2-90B-Vision-Instruct (images); Phi = microsoft/phi-4}
\label{tab:kappa-scores}
\begin{tabular}{llccccc}
\hline
Model & Input & M3(10) & M3(5) & M2 & M1 & M0 \\
\hline
\multirow{2}{*}{Claude} & Image & 0.135 & 0.176 & 0.146 & 0.163 & 0.153 \\
& Text - SKIM & 0.437 & 0.573 & 0.502 & 0.491 & 0.379 \\
\multirow{2}{*}{GPT} & Image & 0.117 & 0.108 & 0.091 & 0.038 & 0.163 \\
& Text - SKIM & \textbf{0.639} & \textbf{0.613} & 0.498 & \textbf{0.650} & 0.257 \\
\multirow{2}{*}{Llama} & Image & 0.245 & 0.237 & 0.192 & 0.118 & 0.131 \\
& Text - SKIM & 0.308 & 0.241 & 0.374 & 0.214 & 0.151 \\
Phi & SText - SKIM & 0.243 & 0.198 & 0.377 & 0.231 & 0.280 \\
\hline
\end{tabular}
\end{table}
\begin{table}[h]
\centering
\caption{Cohen's Kappa scores comparing agreement across different document representations}

\begin{tabular}{lcc}
\hline
\textbf{Model} & \textbf{SKIM} & \textbf{HEAD} \\
\hline
GPT & 0.613 & 0.255 \\
LLaMA & 0.284 & 0.141 \\
\hline
\end{tabular}
\label{tab:document_representation_comparison}
\end{table}

Our agreement analysis between LLM-generated and human relevance judgements revealed significant variation across different model configurations and frameworks (Tables \ref{tab:kappa-scores} \& \ref{tab:document_representation_comparison}). 

Consistently we find that prompts using some domain-specific aspects(\textbf{M1}- \textbf{M3}) do better than prompts that use generic aspects (\textbf{M0}).
The baseline framework performed poorly across all configurations ($\kappa=0.153$-$0.389$), suggesting that educational resource evaluation requires more nuanced frameworks.

The image input modality also performs poorly across various aspect frameworks for the three vision models examined ($\kappa=0.135 - 1.163$).

When deciding how much document content to include in the prompt, more seems to be better (Table \ref{tab:document_representation_comparison}). The \textbf{SKIM} configuration performs better ($\kappa$ increase of 140\% for GPT and 101\% for LLaMA) for both LLMs investigated.
This agrees with previous findings that often, only parts of educational resources are relevant to teachers~\cite{diekema}.

Of the domain-specific aspect frameworks, the participant-derived frameworks (\textbf{M3(5)} and \textbf{M3(10)}) demonstrated the strongest alignment with human judges, achieving Cohen's $\kappa$ scores of 0.639 (PD1-10) and 0.613 (PD1-5).
These scores notably exceed Thomas et al.'s baseline findings for crowd workers ($\kappa=0.24$-$0.52$) and approach their benchmark for controlled lab assessments ($\kappa=0.58$)~\cite{thomas}.

The frameworks from the literature showed variable performance depending on the LLM used. GPT showed better performance for the detailed framework ($\kappa$ = 0.650 for \textbf{M1} vs. 0.498 for \textbf{M2}),
while Claude showed comparable agreement for both.
Both Llama And Phi showed better agreement for the grouped framework (\textbf{M1}).

Across LLMs, the proprietary ones (GPT and Claude with best $\kappa = 0.650$) seem to maintain a sizeable lead over the open source LLMs (Best $\kappa = 0.374 $).

\subsection{System-Level Evaluation Impact}
\begin{table}[t]
\caption{Rank-Biased Overlap Scores for agreement of different rankings produced by LLM judgements with human judgements. RBO scores are calculated with $\phi=0.9$ for query difficulty, $\phi=0.7$ for system rankings. Frameworks per Sec. \ref{sec:frameworks}. Bold values: RBO $\geq$ 0.8; Italic values: $RBO \geq 0.7$.}
\label{tab:rbo-scores}
\begin{tabular}{llcc}
\toprule
& & \multicolumn{2}{c}{RBO Scores} \\
\cmidrule{3-4}
Framework & Metric & Hardest Queries & Best Runs \\
\midrule
& AP & 0.387 & 0.524 \\
M0 & P@10 & \textbf{0.877} & 0.562 \\
& nDCG@10 & 0.372 & 0.562 \\
\midrule
& AP & 0.597 & 0.009 \\
M1 & P@10 & \textbf{0.840} & 0.001 \\
& nDCG@10 & 0.417 & 0.001 \\
\midrule
& AP & 0.641 & 0.562 \\
M2 & P@10 & \textbf{0.886} & 0.562 \\
& nDCG@10 & 0.482 & 0.562 \\
\midrule
& AP & 0.360 & 0.009 \\
M3(10) & P@10 & \textit{0.731} & 0.524 \\
& nDCG@10 & 0.347 & 0.009  \\
\midrule
& AP & 0.505 & 0.045 \\
M3(5) & P@10 & 0.598 & 0.519 \\
& nDCG@10 & 0.476 & 0.519 \\
\bottomrule
\end{tabular}
\end{table}

We use rank-biased overlap (RBO) to measure agreement between LLM and human judgements of ranked systems and queries, we compared four frameworks:
\begin{itemize}
\item M0: Thomas et al.'s~\cite{thomas} two aspects (trustworthiness and content match)
\item M1: Reitsma et al.'s~\cite{reitsma} and Wetzler et al.'s~\cite{wetzler} 12-dimensional framework
\item M2: Higher-level groupings of M1
\item M3: Participant-derived dimensions in 5 and 10-dimension variants
\end{itemize}

\subsubsection{Query Difficulty Assessment}
With P@10, the baseline M0 framework ($RBO = 0.877$), the detailed M1 framework ($RBO = 0.840$), and the grouped M2 framework ($RBO = 0.886$) all showed strong performance in identifying difficult queries, with M2 slightly outperforming the baseline M0.
The participant-derived M3 variants showed weaker performance (M3(10): $RBO = 0.731$, M3(5): $RBO = 0.598$), suggesting that Thomas et al.'s original two aspects were more effective for this task than our participant-identified dimensions.

\subsubsection{System Rankings}
The baseline M0 and hierarchical M2 frameworks achieved identical moderate performance ($RBO = 0.562$) across all metrics.
The detailed M1 framework performed remarkably poorly ($RBO = 0.001$), while the participant-derived M3 variants showed inconsistent results ranging from very poor to moderate ($RBO = 0.009-0.524$).
This suggests no framework improved upon the baseline  for system-level discrimination.

\subsubsection{Comparison to Web Search Evaluation}
Our results show both notable similarities and differences when compared to previous findings. 
For query difficulty rankings, our frameworks consistently outperformed their baseline across all metrics.
While M0 achieved a maximum RBO of 0.48 using MAP@100, our frameworks achieved substantially higher agreement with human judgements, particularly when using P@10 as the metric~\cite{thomas}.
This suggests that LLM judgements are more trustworthy for these kinds of system evaluations only when considering the first page of results rather than the whole ranking.
The baseline \textbf{M0} framework ($RBO=0.877$), detailed \textbf{M1} framework ($RBO=0.840$), grouped \textbf{M2} framework ($RBO=0.886$) and the \textbf{M3(10)} ($RBO=0.731$) framework all demonstrated markedly stronger performance in identifying difficult queries.
However, this did not extend to system rankings.
M0 achieved strong agreement for best runs ($RBO=0.79$ using P@10), our frameworks showed more moderate performance ($RBO=0.562$ for M0 and M2) and in some cases performed poorly ($RBO=0.001$ for M1).

This disparity suggests that while our domain-specific frameworks may be better at capturing the nuanced factors that make educational search queries challenging, they may not translate as effectively to discriminating between retrieval systems.
This could indicate that system-level performance in educational resource retrieval may be less sensitive to domain-specific relevance criteria than query-level difficulty assessment.
The stark contrast between query difficulty and system ranking performance also raises interesting questions about whether different aspects of relevance might be more important for different evaluation tasks in educational resource retrieval.

\subsection{Dimensional Contribution Analysis}

\begin{table}[h]
\centering
\begin{tabular}{llc}
\hline
Framework & Top 3 Dimensions & Total \\
          & (Explained Variance) & Explained \\
\hline
M0 & Match (77.8\%) & 77.8\% \\
 & Trustworthiness (32.5\%) & \\
\hline
M1 & Grade Level (49.0\%) & 64.4\% \\
 & Motivation (45.0\%) & \\
 & Learning Goals (44.9\%) & \\
\hline
M2 & Affective Match (72.3\%) & 89.1\% \\
 & Object Match (64.6\%) & \\
 & Situational Match (59.4\%) & \\
\hline
M3(10) & Implementation (PD7) (57.8\%) & 75.9\% \\
 & Curriculum (PD10) (51.4\%) & \\
 & Content Quality (PD2) (48.6\%) & \\
\hline
M3(5) & Accessibility (PD5) (75.0\%) & 86.3\% \\
 & Grade Level (PD3) (62.1\%) & \\
 & Source Credibility (PD1) (59.3\%) & \\
\hline
\end{tabular}
\caption{Top 3 dimensions in each framework and their contribution to explaining overall relevance judgements. Total explained is the sum of individual contributions, though note that due to correlations between dimensions, this is an upper bound.}
\label{tab:variance}
\end{table}

The variance analysis reveals patterns in how LLMs assess educational resource relevance across different frameworks. 
Notably, the baseline web search framework M0 achieves remarkably high explanatory power (77.8\%) from Match alone), which may indicate that the LLM is falling back on general notions of relevance rather than truly understanding educational-specific criteria. 
While the literature-derived frameworks M1 and M2 provide more nuanced educational dimensions, the LLM's judgements show higher coherence with simpler frameworks - M2's condensed dimensions explain 89.1\% of variance compared to M1's more distributed 64.4\%. 
Similarly, for the teacher-derived frameworks, the LLM's judgements align better with the simpler M3(5) (86.3\%) than M3(10) (75.9\%). 
However, this preference for simpler frameworks might reflect limitations in the LLM's ability to make fine-grained distinctions between multiple specialised dimensions rather than indicating these frameworks' actual utility in practice. 
The high performance of M0 is particularly concerning, as it suggests the LLM may be oversimplifying educational resource relevance to basic topical matching. 

\section{Conclusion}

This study investigated how to effectively use LLMs for evaluating educational resources by addressing two key questions: 
\begin{itemize}
    \item Which combination of relevance dimensions leads to the best agreement with human searchers' judgements?
    \item Do LLM relevance judgements lead to similar query and system rankings as human judgements?
\end{itemize}.
To address these questions, we conducted a user study with teachers and educational experts, developed multiple frameworks for structuring LLM relevance assessments, and evaluated their effectiveness through both direct comparison with human judgements and system-level evaluation metrics.

Our findings revealed three key insights: 
Domain-specific frameworks significantly improve LLM performance, with criteria derived from user study participants achieving particularly strong results ($\kappa$ up to 0.639). This suggests that incorporating user expertise into evaluation frameworks is crucial for professional search contexts.
It raises the discussion of whether it is better to invest energy into gathering localised relevance criteria in small studies rather than relying on large-scale evaluations of global criteria.

LLMs showed varying effectiveness across different evaluation tasks. While they demonstrated high reliability in identifying the hardest queries (RBO 0.713-0.886), their performance in discriminating between individual systems was more moderate (RBO 0.52-0.56). This indicated that LLMs may be more suitable for some evaluation tasks than others. 

The gap between proprietary and open-source LLMs when it comes to professional relevance judgement is quite wide.
This suggests the need for open source benchmarks to improve LLM performance on professional relevance judgement tasks.

Several limitations of our study should be noted. 
Firstly, our evaluation was limited to high school STEM teachers in Germany, and the generalisability to other educational contexts requires further investigation.
Secondly, the performance variation across different input modalities suggests that our understanding of how LLMs process and evaluate different document representations remains incomplete.
Finally, while the investigated frameworks showed promise, we did not yet perform an exhaustive evaluation, e.g. to identify an optimal combination of dimensions of relevance judgement for educational resources.

Future work should focus on (a) the systematic investigation how different dimensions affect assessment reliability across various search tasks; (b) expansion to other educational contexts and subject areas; (c) deeper analysis of LLM behaviour with different document representations.

In conclusion, our findings demonstrate that LLMs can effectively support educational relevance judgements when properly prompted with domain-specific criteria. 
While they may not completely replace human judgement, they show promise as tools for scaling up resource evaluation while maintaining sensitivity to professional requirements.
The success of domain-specific frameworks suggests that careful attention to prompt engineering and domain knowledge incorporation is crucial for implementing LLM-based evaluation in professional search contexts.

\bibliographystyle{ACM-Reference-Format}
\bibliography{refs}

\end{document}